\documentclass{aa}
\usepackage[english]{babel}
\usepackage{graphicx}
\usepackage{amssymb}
\usepackage{hhline}
\usepackage{tablefootnote}
\usepackage{threeparttable}

\begin{document}

  \title{The M\,101 galaxy group as a node in the nearby cosmic filament}
    \titlerunning{M\,101 galaxy group as a node in filament}
    \author{Valentina E. Karachentseva\inst{\ref{inst1}},
    Igor D. Karachentsev\inst{\ref{inst2}},
    Elena I. Kaisina\inst{\ref{inst2}} \thanks {contact e-mail: kaisina.elena@gmail.com},
    Serafim S. Kaisin\inst{\ref{inst2}}}
    
    \authorrunning{V.\,Karachentseva et al.}

    \institute{Main Astronomical Observatory, National Academy of Sciences of Ukraine, Kiev, 03143, Ukraine \label{inst1}
    \and Special Astrophysical Observatory of the Russian Academy of Sciences, N.Arkhyz, KChR, 369167, Russia \label{inst2}}
    
    \abstract {We performed a search for faint low surface brightness dwarf 
galaxies around the major spiral galaxy M\,101 and in the large 
rectangular area within SGL = [30 -- 80]$^{\circ}$, SGB =[10 -- 37]$^{\circ}$ spanning a 
chain of galaxies: M\,63, M\,51, M\,101, and NGC\,6503, based on the 
data from DESI Legacy Imaging Surveys. Six new supposed 
dwarf members of the complex were discovered. We present a list of 
25 prospective members of the M\,101 group and estimate the total 
mass and the total-mass-to-$K$-band luminosity ratio of the group as $(1.02\pm0.42)\times10^{12}~M_{\odot}$ 
and  $(16.0\pm6.5)~M_{\odot}/L_{\odot}$, respectively. We 
notice that the average dark mass-to-luminosity ratio in the groups 
around M\,63, M\,51, and M\,101 is $(12\pm4)M_{\odot}/L_{\odot}$ that almost an order 
of magnitude lower than the global cosmic ratio, $(102\pm5)M_{\odot}/L_{\odot}$. }


\maketitle

 \section{Introduction}
 The filaments as the extending elongated structures, which consist of 
many galaxies, are one of main elements of the large-scale structure of 
the Universe, together with walls (sheets, pancakes), rich clusters, and 
cosmic voids \citep{Joeveer1978,Bond1995,Cautun2014}.
The reality of the cosmic filaments is confirmed obviously by many N-body simulations made in
the standard $\Lambda$CDM cosmological model \citep{Aragon2008,Chen2015,Libeskind2018,Libeskind2020}. Characteristic linear dimensions, population, and a degree of 
``straightness'' of the filaments remain not yet quite fixed, because they 
depend strongly on the criteria, by which the cosmic filaments have been 
highlighted. The most rigorous definition of a filament as one-dimensional sequence 
(a chain) of galaxies was proposed by Tempel et al. 
(2014). They did publish a  catalog of filaments based on the 
Sloan Digital Sky Survey (SDSS) Data Release 8 (DR8) \citep{York2000, Aihara2011}. Linear dimensions of 
these filaments are equal to $\sim(3 - 60)$~Mpc. Here is still not clear 
understanding of whether the space between the members of the filament is 
filled with neutral or hot gas and what is the proportion of dark-to-light 
matter in these structures. 

The answer to these important questions can be obtained by studying 
the closest objects of this type. 
Within the Local supercluster, the best-known filament is Virgo Southern 
Extension  (VSE) \citep{Tully1984}, whose length reaches 
approximately 6~Mpc. 
Using the data on distances and radial velocities of galaxies in VSE, 
\citet{Kashibadze2020} showed that galaxies of this filament are moving 
to the Virgo cluster center, replenishing its population. \citet{Kim2016} 
and \citet{Castignani2022} marked around the Virgo cluster a dozen of other filament 
structures of a similar length. However, the population and kinematics of 
these filaments are still almost not investigated. 

 In the nearest Local volume limited by a distance of $D\sim10$~Mpc around the
Milky Way, several candidates to linear structures have been marked. \citet{Jerjen1998}, 
\citet{Karachentsev2003}, and \citet{Pruzhinskaya2020} suggested that nearby 
galaxies in Sculptor constellation, i.e. NGC\,55, NGC\,300, NGC\,247, NGC\,253, NGC\,24, NGC\,45 and 
their satellites form a diffuse chain which is elongated in the radial 
direction in the distance interval from 2 to 7~Mpc. According to \citet{Sandage1974},
the northern sky bright galaxies: IC\,342, NGC\,2403, M\,81, NGC\,4236 with their satellites
can be members of a chain located approximately in the picture plane at a 
distance of 3 -- 4~Mpc from us. \citet{Muller2017} and \citet{Karachentsev2020} supposed 
that the galaxy groups around massive spiral galaxies M\,63 (NGC\,5055), M\,51 (NGC\,5194) 
and M\,101 (NGC\,5457) are part of a filament that extends towards the Local Void, 
including possibly the NGC\,6503 galaxy too.

 \citet{Karachentsev2020} and \citet{Karachentsev2021}  reviewed 
recently the structure and kinematics of the galaxy groups around M\,63 and M\,51. 
In this work, we discuss new data about the M\,101 group population in the context 
of its membership in the M\,63/M\,51/M\,101 filament. The proximity of this chain of groups 
allows make a clearer portrait of the cosmic filament, while it is easier to 
measure distances to nearby galaxies and to exclude projected foreground and 
background objects.

 The paper is outlined as follows. We begin in Section 2 by describing 
the searches for supposed new low surface brightness dwarf galaxies 
and their photometry. Using the radial velocities and projected separations
for eight M\,101 group galaxies, we estimate the total (orbital) mass of the group. In 
Section 3, we discuss the structure and population of a chain of galaxy 
groups, which includes galaxies  M\,101, M\,51, M\,63, and possibly NGC\,6503, together with 
their satellites. We calculate  the dark-to-luminous mass matter in the 
filament groups, which is almost an order of magnitude lower than the 
global cosmic ratio. In Section 4, we give our primary results, focusing on 
the general size of the galaxy filament and briefly summarize our 
results. It is also noticed that the environment of the filament is extremely poor, 
partly belonging  to the Local Void.

In Appendix \ref{appendix}, we present a list of 62 supposed members of the filament as well as 
10 galaxies in the immediate environment of the filament. In all calculations we use the Hubble 
parameter, $H_0 = 73$~km~s$^{-1}$~Mpc$^{-1}$.

\section{Search for new dwarf galaxies in the M 101 group}

For a long time, it was believed that only a few late-type dwarf 
galaxies, i.e. NGC\,5474, NGC\,5477, NGC\,5585, and Holmberg\,IV 
are associated with massive spiral galaxy M\,101. 
In the last decade, searches have been undertaken for fainter satellites of M\,101. 
\citet{Merritt2014} and \citet{Karachentsev2015} performed a survey of the 
surroundings of M\,101 using long exposures on small telescopes with 
a large field of view. As a result, 10 low surface brightness 
candidates to the satellites of M\,101 were discovered. \citet{Bennet2017} 
did repeat the survey of the virial zone around M\,101, obtaining  
deeper images of this region at the 3.6-m CFHT telescope. 
On these images, made with high resolution, they detected 38 dwarf spheroidal galaxies with apparent magnitudes ranging between 19 and 21 mag in the B-band, and angular diameters of  $\sim(10 - 20)\arcsec$. However, it later turned 
out that the most part of them are associated with distant galaxy 
groups NGC\,5422 and NGC\,5485 located behind M\,101 at a distance of
$D\simeq29$~Mpc \citep{Karachentsev2019}. \citet{Muller2017} 
undertook a search for dwarf galaxies outside the virial zone of M\,101
using the data of SDSS DR 11-12 \citep{Alam2015}.  
As a result, they found six new candidates to the M\,101 satellites.

At the Hubble Space Telescope (HST) with the ACS camera, there 
were obtained images in filters F606W and F814W for two dozen 
supposed satellites of M\,101 (the programs GO\,13682, PI van Dokkum and GO\, 
14796, PI Crnojevich). Based on these images, distances to 
a number of dwarfs resolved into stars were determined by the luminosity of
the tip of the red giant branch ($=$TRGB) \citep{Merritt2016, Danieli2017, Karachentsev2019}. Using the images obtained with CFHT, \citet{Carlsten2022} performed a surface photometry of many dwarf galaxies in the M\,101 
virial zone and determined  their distances by the surface brightness fluctuation (sbf) method.

 The galaxy group around M\,101 is located in the area of DESI Legacy Imaging Surveys \citep{Dey2019}. This survey  is deeper than the SDSS sky survey by about 1.5 mag. We use the DESI Legacy Imaging Surveys data to search for new dwarf 
satellites of M\,101. Our search region covered an area with a radius of 6$^{\circ}$ 
around M\,101, which is almost 4 times the virial radius of the 
group. Emphasis was placed on finding objects of low and very low 
surface brightness. As a result, we found 5 candidates for M\,101 
peripheral satellites. The data of them are presented in Table \ref{table1}.
Its columns indicate: the galaxy name; equatorial coordinates in degrees; supergalactic coordinates in degrees; an angular diameter of the galaxy in arc minutes, maximum visible in the DESI Legacy Imaging Surveys; the apparent axis ratio; 
morphological type; the total $g$ and $r$ band mag and the total
$B$ mag estimated via the relation $B=g+0.313(g-r)+0.227$, 
recommened by Lupton\footnote{https://www.sdss3.org/dr10/algorithms/ sdssUBVRITransform.php\#Lupton2005}.  The last raw of Table~1 contains the data on a new dwarf galaxy found by us slightly away from 
the M\,63 and M\,51 groups. 

\begin{table*}
\centering
\caption{\label{table1}New candidates for M 101 satellites}
\begin{tabular}{lccccclrrr}
\hline
 Name       &RA (2000.0) DEC & SGL & SGB &    $a$    &  $b/a$  &  Type &   $g$&  $r$  &  $B$ \\
 \hline
            &   degree  & degree & degree  & arcmin &         &         &    mag& mag& mag\\  
\hline
 KK 207 \tablefootmark{a} &203.357+56.500 & 61.43 &  18.32  &   0.97   & 0.70  & dIrr  &   18.41  & 18.07 &  18.75\\
 Dw1348+60  &207.024+60.067  & 57.61 & 20.18 &  0.83   & 0.75  & dTr   &    19.45 &  18.67&   19.92  \\
 Dw1351+50  &207.996+50.248  & 68.07 & 21.08 &  0.50   & 0.82  & dIrr  &    20.10  & 19.56&   20.50\\
 Dw1358+52  &209.531+52.918  & 65.18 & 21.95 &  0.72   & 0.92  & dSph  &   19.86   &19.03 &  20.35 \\
 Dw1409+51  &212.304+51.225  & 66.89 & 23.76 &  1.70   & 0.58  & dIrr  &   18.60  & 18.15 &  18.97 \\
\hline
 Dw1341+42  &205.495+42.069  & 76.72 & 19.08 &  0.42   & 0.77  & dIrr  &   19.80  & 19.75 &  20.04 \\
\hline
\end{tabular}
\tablefoot{
\tablefoottext{a}{Re-discovered by us. See note to this objects.}
}
\end{table*}

The apparent $g$ and $r$ magnitudes were obtained from  surface photometry of galaxies in the DESI Legacy Imaging Surveys DR 9.
We measured fluxes of these galaxies in $g$- and $r$-bands using
the standard methods of processing in the ESO-MIDAS software package. 
Foreground stars were removed from the frames by fitting a second-degree surface in circular pixel area. The sky background on the image was approximated by a tilted
plane, created from a 2-dimensional polynomial using the least-squares method (fit/flat sky). The mean uncertainty introduced by the inaccuracy of the sky background determination is $\leq0.12$ mag, being primarily caused by the background variations across the frame. A part of our sample is comprised 
of galaxies of extremely low surface brightness.
The objects are characterized by an irregular, clumpy structure. For this
reason, we did not approximate the galaxies by ellipses, but used circular apertures. The center of each galaxy was determined interactively.
To measure fluxes in $g$- and $r$-bands the integrated photometry was performed in increasing circular apertures from a pre chosen center to the faint outskirts
of the galaxies. The total flux was then estimated as the asymptotic value of the obtained radial growth curve. The uncertainties of the total flux determination
were $0.06$ mag in $g$- and $0.08$ mag in $r$-bands. We obtained the apparent magnitudes according to the relation $m = 22.5 - 2.5\log {\rm (flux)}$\footnote{https://www.legacysurvey.org/dr10/description/}. 

\begin{figure}
\centering  
\resizebox{\hsize}{!}{\includegraphics{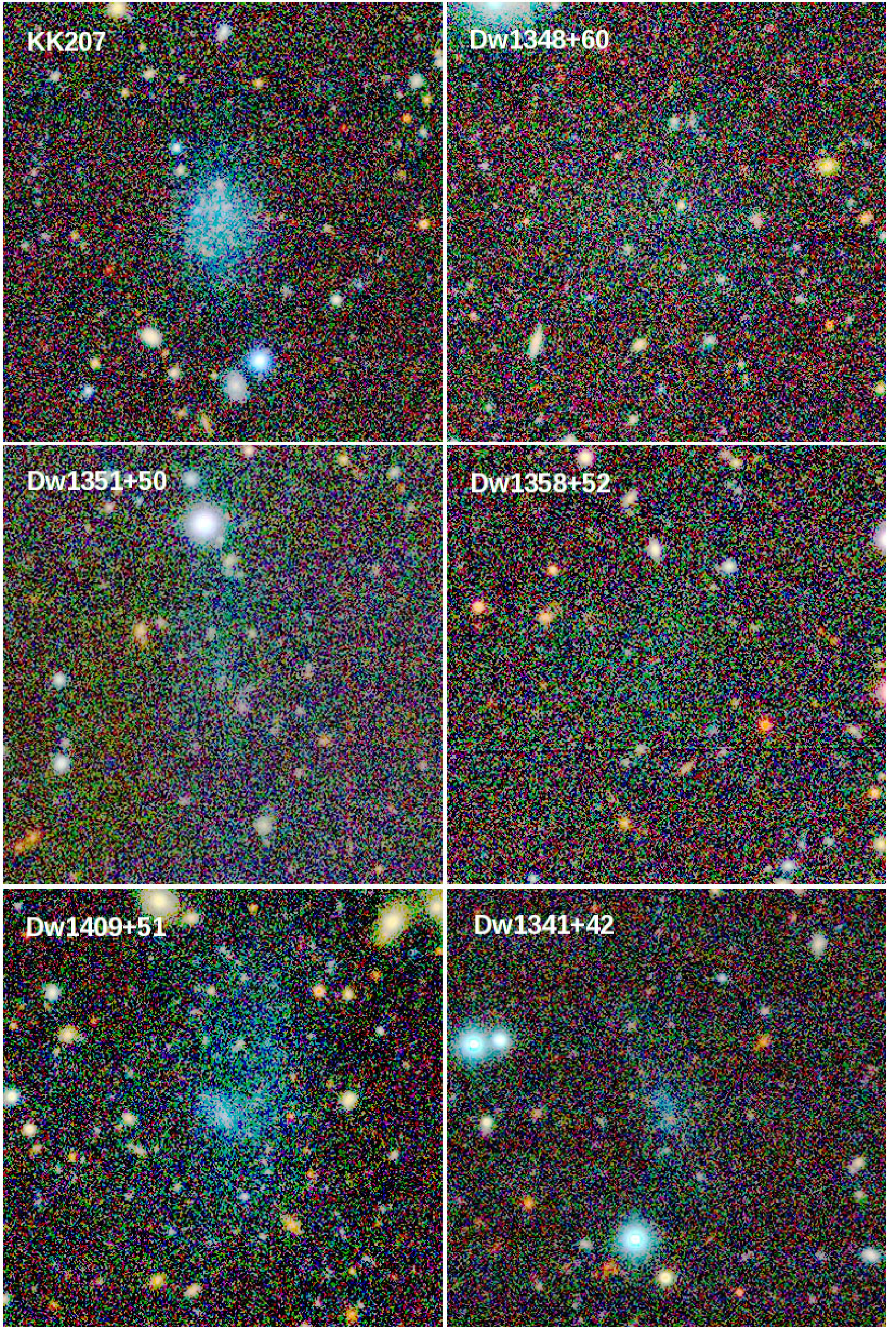}}
\caption{Images of 6 new dwarf galaxies from the DESI Legacy Imaging Surveys found in the M\,63/M\,51/M\,101/NGC\,6503 filament. Each image size is  $2^{\prime}\times 2^{\prime}$. North is to the top, East is to the left.}
\label{figure1}
\end{figure}

The mosaic of images of these galaxies, taken from the DESI Legacy Imaging Surveys, is given in Fig.\ref{figure1}. One side of an image corresponds to 2$\arcmin$. North is in the top, east is in the left.

 Below, we note some features of the discovered galaxies.

  {\em KK\,207.} This dwarf galaxy of irregular type was found earlier 
\citep{Karachentseva1998} on the Second Palomar Observatory Sky Survey, but not assigned to the M 101 group. Its structure 
seems as granulated one on the DESI Legacy Imaging Surveys. The galaxy is also detected in the 
GALEX survey \citep{Martin2005} and has the apparent magnitude of $m({\rm FUV}) = 20.68$ mag.

   {\em Dw\,1348+60.}  The dwarf galaxy of spheroidal (dSph) or transition (Tr) type of very low
surface brightness. Probably it is a companion of a distant galaxy NGC\,5322 with  
$V_{\rm LG} =1932$~km~s$^{-1}$ and $D=31.9$~Mpc
(sbf), which has a number other small dSph
satellites. However, at a distance of 31.9~Mpc, the galaxy Dw\,1348+60 would have
a linear diameter of 8~kpc, which is quite unusual for a dwarf galaxy with such a
low surface brightness. On the other hand, Dw\,1348+60 is located on far outskirts
of the M\,101 group, which raises the question of the cause of suppressed star-formation
in this galaxy.

  {\em Dw\,1351+50.} The dwarf galaxy of irregular- or transition-type 
of very low surface brightness, which is barely visible in the SDSS survey. 

  {\em Dw\,1358+52.} This dwarf spheroidal galaxy of very low surface 
brightness is placed in the survey zone of \citet{Carlsten2022} but is 
not noticed by them.

  {\em Dw\,1409+51.} This is a dwarf irregular galaxy of low surface brightness with a
blue central knot  having in GALEX $m({\rm NUV}) = 20.09$ mag. A distant Sd-galaxy 
UGC\,9050 is lying at $8\arcmin$ to West from it. This neighboring galaxy has
a heliocentric velocity of $V_{h} = 2001$~km~s$^{-1}$ and a distance estimate of 32~Mpc.
Both galaxies can form a physical pair. But in this case the linear diameter of
Dw\,1409+51 could reach about 16 kpc, which is atypical for such a dwarf object.\footnote {In a recent paper, C. Fielder et al. (arXiv:2306.06164) studied Dw1409+51 = UGC 9050-Dw1 with Hubble Space Telescope and Very Large Array. They found the object to be a background ultra-diffuse galaxy with a heliocentric velocity of $1952$~km~s$^{-1}$.}

  {\em Dw\,1341+42.} Isolated dwarf irregular galaxy with slightly 
granulated structure and GALEX mag of $m({\rm NUV}) = 22.14$ mag.

 \begin{table*}
\caption{\label{table2}Galaxies associated with M\,101}
\begin{tabular}{lcclcrlrcc} \hline

  Name        &        RA (2000.0)  DEC&    $D$ &   meth &$V_{\rm LG}$ &  $\log L_K$ &  Type     & $R_p$    &    Ref     &    $M_{\rm orb}$\\ \hline
  (1)         &          (2)           &    (3) &   (4)  &   (5)  &  (6)    &   (7)     &   (8)    &     (9)    &     (10)\\ \hline
 KK 207       &     133325.7+563000    &  6.95  &   mem  &    -   &      6.47  &    dIrr   &     575  &      (1)   &      -  \\
 dw1343+58    &     134307.0+581340    &  5.59  &   TRGB &   338  &   7.46  &    BCD    &     582  &      (2)   &     1.10 \\
 Dw1348+60    &     134805.8+600401    &  6.95  &   mem  &   -    &    6.75   &    dTr    &     737  &      (1)   &       - \\
 dw1350+5441  &     135058.3+544120    &  6.27  &   sbf  &   -    &   7.54  &    dSph   &     220  &      (3)   &      - \\
 Dw1351+50    &     135159.0+501453    &  6.95  &   mem  &   -    &   5.75   &    dTr    &     545  &      (1)   &      - \\
 Holmberg IV  &     135445.1+535417    &  7.24  &   TRGB &   272  &   8.62  &    Sm     &     163  &      (2)   &     2.16\\
 GBT1355+54   &     135450.6+543850    &  6.95  &   mem  &   345  &   7.10  &    HIcld  &     152  &      (4)   &     0.21\\
 dw1355+51    &     135511.0+515429    &  6.95  &   mem  &   -    &   7.07  &    dSph   &     334  &      (5)   &      - \\
 M101Dw9      &     135544.6+550845    &  7.73  &   TRGB &   -    &   6.26  &    dSph   &     166  &      (6)   &      - \\
 UGC 8882     &     135714.6+540603    &  6.95  &   sbf  &   482  &   7.51  &    dEn    &     110  &      (3)   &     1.40\\
 Dw1358+52    &     135807.4+525505    &  6.95  &   mem  &    -   &   6.52   &    dSph   &     199  &      (1)   &       -\\
 M101-df3     &     140305.7+533656    &  6.52  &   TRGB &   -    &   7.27  &    dSph   &      97  &      (7)   &       - \\
 M 101        &     140312.8+542102    &  6.95  &   TRGB &   378  &   10.79 &    Scd    &       0  &      (2)   &      -  \\
 M101-df1     &     140345.0+535640    &  6.37  &   TRGB &   -    &   6.17   &    dTr    &      53  &      (7)   &       -  \\
 NGC 5474     &     140502.1+533947    &  6.98  &   TRGB &   424  &   9.21  &    Sm     &      96  &      (2)   &      0.24\\
 NGC 5477     &     140533.1+542739    &  6.76  &   TRGB &   451  &   8.26  &    Im     &      44  &      (2)   &      0.28\\
 M101-DwA     &     140650.2+534432    &  6.65  &   TRGB &     -  &   6.68  &    dSph   &     102  &      (6)   &       - \\
 M101Dw7      &     140721.0+550351    &  6.95  &   mem  &    -   &   6.14  &    dSph   &     115  &      (8)   &       - \\
 M101-df2     &     140837.5+541931    &  6.87  &   TRGB &    -   &   6.74  &    dSph   &      96  &      (7)   &        - \\ 
 Dw1409+51    &     140913.0+511330    &  6.95  &   mem  &    -   &   6.38   &    dIrr   &     403  &      (1)   &        - \\
 dw1412+56    &     141211.0+560831    &  6.95  &   mem  &    -   &   6.79  &    dSph   &     272  &      (5)   &        -  \\
 dw1416+57    &     141659.0+575439    &  6.95  &   mem  &    -   &   6.95  &    dSph   &     499  &      (5)   &        - \\
 NGC 5585     &     141948.3+564349    &  7.00  &   TRGB &   457  &   9.21  &    Sdm    &     410  &      (2)   &      3.02 \\
 DDO 194      &     143524.6+571524    &  5.81  &   TRGB &   381  &   8.09  &    Sm     &     661  &      (2)   &      0.01 \\
 dw1446+58    &     144600.0+583404    &  6.95  &   mem  &     -  &   6.49  &    dIrr   &     911  &      (5)   &        - \\
 \hline
 \end{tabular}
\tablebib{(1) this work; (2) \citet{Anand2021}; (3) \citet{Carlsten2022};
 (4) \citet{Mihos2012}; (5) \citet{Muller2017}; (6) \citet{Karachentsev2019};
 (7) \citet{Danieli2017}; (8) \citet{Bennet2017}.} 
\end{table*}

At present, a list of members of the M\,101 group includes 25 candidates. A 
summary of them is presented in Table \ref{table2}. Its columns contain: (1)~--- galaxy name; (2)~--- equatorial coordinates; (3)~--- distance to galaxy in Mpc; (4)~--- method, by which the galaxy distance estimation is made: ``TRGB''~--- by the luminosity of the tip of red giants branch, ``sbf''~---  by the surface brightness fluctuations, 
``mem''~--- by supposed group membership; (5)~--- radial velocity of galaxy in km~s$^{-1}$ 
relative to the centroid of the Local Group; (6)~--- logarithm of the 
integral luminosity of  galaxy in the $K$-band , in units of the
luminosity of the Sun, taken from Updated Nearby Galaxy Catalog ($=$UNGC) \citep{Karachentsev2013}; (7)~--- morphological type taken from UNGC;
(8)~--- projected separation of the galaxy from M\,101 in kpc assuming 
that all objects are at the same distance as M\,101; (9)~--- link  to source of 
the distance data; (10)~--- orbital estimate of the total mass of the group 
by projected separation, $R_p$, and radial velocity difference, $\Delta V$, of each satellite relative to M\,101: $M_{\rm orb}= (16/\pi\times G) R_p \Delta V^2$
\citep{Karachentsev2021}, where $G$ is the gravitational constant and 
the mass is expressed in units of $M_{\odot}$. In the list of group members we included  
also the intergalactic hydrogen  cloud GBT\,1355+54  found by \citet{Mihos2012}.

 According to these data, the average value of the orbital 
estimate of the total mass of the M\,101 group is  $\langle M_{\rm orb}\rangle  =(1.05\pm0.42)\times10^{12}~M_{\odot}$. 
With the integral luminosity of the group  $\Sigma L_K=6.56\times 10^{10}~L_{\odot}$ the total 
mass-to-luminosity ratio, $\langle M_{\rm orb}\rangle/\Sigma L_K$, is equal to  $(16.0\pm6.5)~M_{\odot}/L_{\odot}$.
This  quantity is typical for the groups where the late-type spiral galaxy dominates.

 It should be noted that the membership of some peripheral 
galaxies in the M\,101 group is questionable. So, the dwarf galaxies of 
very low surface brightness: Dw\,1348+60 and Dw\,1409+51 may be satellites of 
distant galaxies NGC\,5322 and UGC\,9050 at a distance of about 32~Mpc.
Besides, dwarf galaxies dw\,1446+58 and dw\,1416+57, found  by \citet{Muller2017}, 
look in the DESI Legacy Imaging Surveys as background galaxies. Being without measured radial velocities, these 4 galaxies do not participate in the orbital mass estimation.

\section{The chain of groups M\,63/M\,51/M\,101/NGC\,6503} 

The basic parameters of these four groups are presented in Table \ref{table3}, 
whose columns contain: (1)~--- the main galaxy name; (2)~--- its supergalactic coordinates; 
(3, 4)~---  distance (Mpc) and radial velocity (km~s$^{-1}$)  of the host galaxy; 
(5)~---  number of the group member candidates; (6)~--- number of satellites with a 
measured radial velocity; (7)~--- the total (orbital) mass in $M_{\odot}$ units; 
(8)~---  virial radius of the group (kpc) estimated from \citet{Tully2015} relation
 ($R_{\rm vir}/215$~kpc) $= (M_T/10^{12}~M_{\odot})^{1/3}$ ; (9)~--- the total mass-to-luminosity 
ratio. For the groups of galaxies M\,63 and M\,51 their mass estimates are taken 
from \citet{Karachentsev2021} , and for\,NGC 6503 the mass estimate 
is made by only one satellite \citep{Karachentsev2022}.
\begin{table}
\caption{ \label{table3}Galaxy groups in the M 101 filament}
\resizebox{\hsize}{!} {
\begin{tabular}{lcccrclcc}
\hline

 Name      &   SGL  SGB    &     $D$ &   $V_{\rm LG}$ &    $n$  &   $n_v$ &   $\log(M_T)$ &   $R_{\rm vir}$ &$M_T/L_K$   \\
\hline
  (1)         &          (2)           &    (3) &   (4)  &   (5)  &  (6)    &   (7)     &   (8)    &     (9)    \\ \hline
 M 63      &  76.20+14.25  &  9.04   &  562   &  21   &   7   &   11.71    &    172  &        5$\pm$2 \\
 M 51      &  71.12+17.33  &  8.40   &  538   &  16   &   5   &   12.13    &    237  &       10$\pm$8 \\
 M 101     &  63.58+22.61  &  6.95   &  378   &  25   &   8   &   12.02    &    220  &       16$\pm$6 \\
 N 6503    &  33.14+34.63  &  6.25   &  309   &   2   &   1   &   11.5     &    126  &        40    \\
 \hline
 \end{tabular}}
 \end{table}

\begin{figure}
\centering
\resizebox{\hsize}{!}{\includegraphics{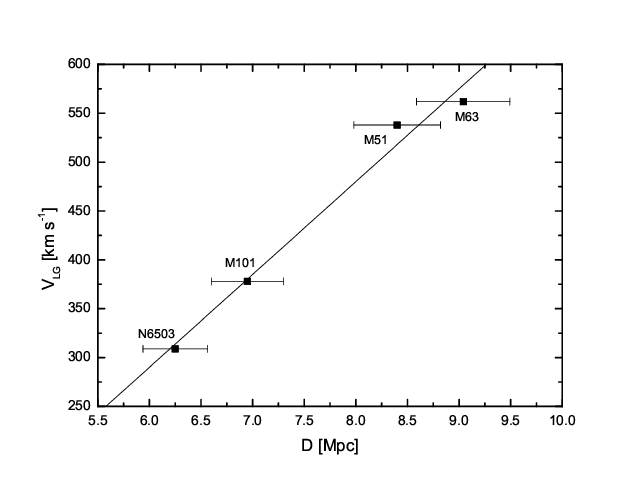}}
\caption{Relation between radial velocities and distances of the principal galaxies in the M\,101 filament.
            The horizontal bars correspond to 5-percent errors in the galaxy distances. The oblique
            line indicates the linear regression with a slope of  $H = (95\pm7)$~km~s$^{-1}$Mpc$^{-1}$.}
\label{figure2}
\end{figure}

As can be seen from these data, the galaxy NGC\,6503 with its 
satellite KK\,242 is quite far from other members of the filament. 
Nevertheless, its distance and radial velocity fit well with the general 
relation between $V_{LG}$ and $D$ (see Fig. \ref{figure2}). The linear regression in the Fig. \ref{figure2} 
passes through each a 5\%~--- distance  error interval for all the galaxies. 
The slope of the regression line in the Fig. \ref{figure2} corresponds formally to the Hubble parameter  $H=(95\pm7)$\,km~s$^{-1}$Mpc$^{-1}$. The large value of the Hubble parameter is explained in the Numerical Action Methods model \citep{Shaya2017}, which describes the local velocity field taking into account the expansion of the Local Void and the Virgo-centric flow. According to a "Distance - Velocity" diagram in the Extragalactic Distance Database\footnote{http://edd.ifa.hawaii.edu/NAMcalculator} \citep{Kourkchi2020}, the galaxy M\,63 at the far end of the filament has a peculiar radial velocity of $-50$~km~s$^{-1}$ relative to the unperturbed Hubble flow with $H=73$~km~s$^{-1}$Mpc$^{-1}$, whereas, the galaxy NGC\,6503 at the close end of the chain has a peculiar velocity of $-115$~km~s$^{-1}$. This difference in peculiar velocities results in a steeper observed slope, $H=96$~km~s$^{-1}$Mpc$^{-1}$ instead of $73$~km~s$^{-1}$Mpc$^{-1}$.

\begin{figure*}
\centering
\includegraphics[height=9cm]{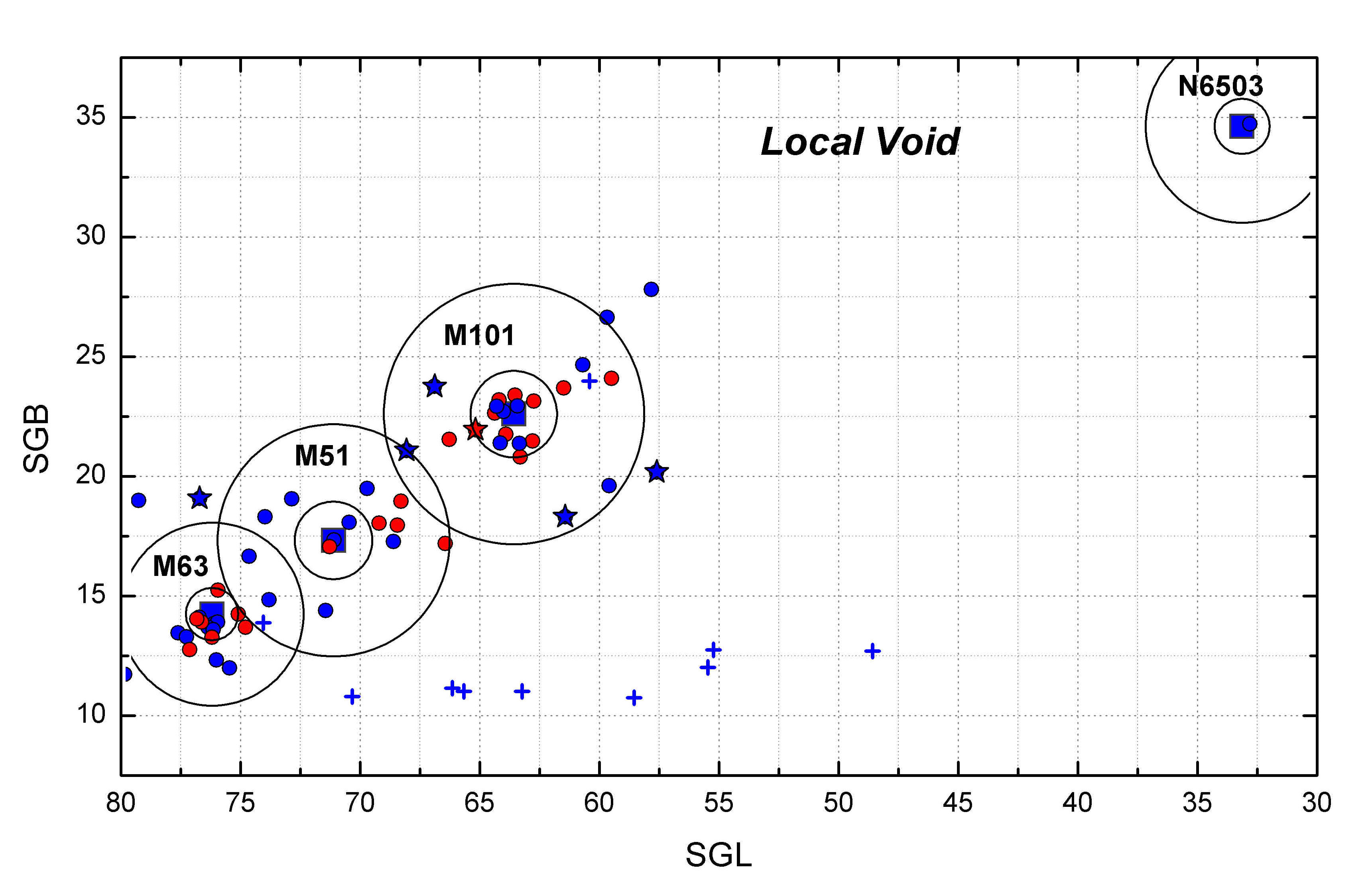}
\caption{Survey area of about 1200 square degrees in the M\,63/M\,51/M\,101/NGC\,6503 filament region.
            The filled circles identify supposed satellites of the principal galaxies in the chain of groups,
            which are shown as squares. Crosses identify other galaxies with distances from 5 to 10 Mpc,
            not assosiated with the filament. The dwarf galaxies discovered by us are shown by asteriscs.
            The red symbols correspond to early-type galaxies ($T < 0$), while the blue symbols represent
            late-type ones. Small and large circles around the principal galaxies indicate their virial
            radius, $R_{\rm vir}$, and the zero-velocity radius, $R_0$.}
\label{figure3}
\end{figure*}

A general shape of the chain of considered groups in supergalactic coordinates 
is shown in Fig. \ref{figure3}. The early-type  and late-type galaxies depicted by red and blue 
symbols, respectively.  The host galaxies in groups are indicated by squares. 
Small and large circles around them correspond to the virial radius of group, $R_{\rm vir}$, 
and the radius of zero-velocity sphere, $R_0 \simeq 3.5~R_{\rm vir}$,  which separates the 
volume of the group from the general expanding cosmic space. Six  new dwarf galaxies
found by us are indicated by asterisks. They can be identified by their supergalactic coordinates presented in Table 1.

 In addition to the alleged members of specified groups, we show by 
crosses in Fig.~\ref{figure3} other galaxies from the UNGC catalog with their 
distance estimate in the interval (5 -- 10)~Mpc. Almost all of them (9 
out of 10) lie near the Local supergalactic plane (SGB $< 15^{\circ}$).

 General summary of data on 72  galaxies in the region SGL = $[30^{\circ}- 80^{\circ}]$,
SGB = $[10^{\circ} - 37^{\circ}]$ with the distance estimates $D = (5-10)$~Mpc is 
presented in Appendix~\ref{appendix}. The columns of this table contain: (1)~--- the galaxy name; 
(2,3)~---  supergalactic coordinates in degrees; (4)~--- $K$-band integral luminosity in the Sun
luminosity units, taken from UNGC; (5)~--- morphological type from UNGC in de 
Vaucouleurs digital  scale; (6)~---  radial velocity relative to
the Local group centroid, in km~s$^{-1}$; (7)~--- distance in Mpc; (8) method, 
by which the distance estimation is made; here ``SN'' denotes  the distance 
estimated from the luminosity of supernovae, ``TF'' is the estimate by 
Tully-Fisher relation between the galaxy rotation amplitude and  luminosity, 
``NAM'' is a kinematic estimate taking into account the local velocity field 
\citep{Shaya2017},  and ``txt'' is the rough estimate by a galaxy texture.
The galaxies in each group are ranked according to their SGL. In Appendix \ref{appendix} and in Table~\ref{table2}, we distinguish between the dwarf galaxies that we discovered and known from the literature as "Dw" and "dw", respectively.

 We examined in DESI Legacy Imaging Surveys all sky region outlined in 
Fig.~\ref{figure3} of a total area of about 1200 square degrees and 
found only one dwarf galaxy with supposed distance of $D < 10$~Mpc. 
The galaxy is shown in last row of Table~\ref{table1}. This fact tell us that the main body 
of filament M\,63/M\,51/M\,101/ NGC\,6503 is indeed surrounded by almost
empty space, being  part of the Local Void, which extends towards the northern supergalactic pole.

 The M\,63 group as the farthest side of filament from us is the closest one
to the Local Sheet plane. Wherein, we did not find any massive 
group in the Local Sheet itself which could serve as a southward 
continuation of the filament under consideration.

 As seen from Fig.~\ref{figure3}, the zones of gravitational influence of the 
massive spiral galaxies M\,63, M\,51, and M\,101, outlined by the radius 
$R_0$, intersect with each other. This means that in the 
future these groups will merge into a single system. The galaxy NGC\, 
6503 with its companion stand apart from the main body of the 
filament.

The dwarf spheroidal galaxies without obvious sings of star 
formation are located inside the $R_0$ zones, demonstrating the 
well-known segregation effect of the early type galaxies ($T < 0$)
against the late type ones in dependence on the density of 
environment.

 From the data of Table~\ref{table3} follows that the average weighted 
ratio of the total mass-to-$K$-band luminosity for the M\,63, M\,51, 
and M\,101 groups is $(12\pm4)~M_{\odot}/L_{\odot}$. According to \citet{Driver2012}, 
the average global density of the K- luminosity is equal to $(4.3\pm0.2)\times 10^8~L_{\odot}$/Mpc$^3$
at  $H_0=73$~km~s$^{-1}$Mpc$^{-1}$. At the value of the critical cosmic density $1.46\times 10^{11} M_{\odot}$~Mpc$^{-3}$ and 
the parameter  $\Omega_m = 0.3$, the ratio of  the global density of dark  
matter to the global density of $K$-luminosity is $(102\pm5)~M_{\odot}/L_{\odot}$.
Therefore, the ratio of dark-to-luminous matter in the M\,63/ M\,51/ M\,101 
filament groups turns out to be  8.5 times less than the global cosmic ratio. 
In order for this chain of groups not to look like a ``dent'' in the general cosmological 
field, it is necessary to assume that the bulk of dark matter of the filament 
is distributed beyond the virial radii of these three groups. In such a case, the estimates of virial radius from the Tully's relation would be rather uncertain. The reason for the low dark-to-stellar mass ratio in the groups under consideration remains unclear to us. Possibly, the dark mass deficit of the filament groups can be caused by their proximity to the Local Void. A special deep survey of this filament and its vicinity in the 21 cm line could reveal new small dwarfs and HI- clouds in the system, thereby increasing the statistical reliability of estimates of the total mass of the groups. 

\section{Concluding remarks on the filament}

The extended galaxy chains, i.e. filaments, are a common 
configuration in the large-scale structure of Universe. According to \citet{Tempel2014}, about 35 -- 40\% of all galaxies enter into cosmic filaments.
We have considered one of their nearest representatives, which is inside the 
Local Volume. Main elements of this filament are the groups around the massive 
spiral galaxies M\,63, M\,51, and M\,101, which are in contact with each 
other. The spiral galaxy NGC\,6503 with its satellite KK\,242 may serve 
as a continuation of this chain. The M\,63, M\,51, M\,101, and NGC\, 
6503 radial velocities as well as their distances from the observer
smoothly change along the body of the filament. The far end of this 
chain at the distance of 9~Mpc is adjacent to the Local Sheet plane, while 
the near end at the distance of 6~Mpc is embedded in the Local Void volume. 
A total filament length, taking into account NGC\,6503, is $\sim60^{\circ}$ or 8~Mpc.

 Using the data of  DESI Legacy Imaging Surveys we have undertaken the 
searches for new dwarf galaxies in a wide region of $50^{\circ}\times 27 ^{\circ}$
designated in Fig.~\ref{figure3}. In the M\,101 group itself, 5 new candidates for 
satellites of the host galaxy were discovered, and only one dwarf galaxy was 
found outside the filament body. The list of 72 galaxies in this region with 
the distance estimates from 5~Mpc to 10~Mpc is presented in Appendix~\ref{appendix}. 
Of these, only 10 galaxies are located outside the filament body, and 
most of them, 9 out of 10, have a low supergalactic latitude, tending 
towards the Local Sheet.

 The proximity of this cosmic filament made it possible for the 
first time to estimate the ratio of dark to light matter in it, 
$M_T/L_K$. The average weighted ratio $\langle M_T/L_K\rangle\simeq(12\pm4)~M_{\odot}/L_{\odot}$
for the filament groups turned out to be almost an order of magnitude less
than the global ratio $(102\pm5)~M_{\odot}/L_{\odot}$ in the Standard Model $\Lambda$CDM. 
The reason for this paradox deserves careful analysis.

 Despite great efforts made in the last quarter of the century to 
measure accurate distances with the Hubble Space Telescope, more 
than 60\% galaxies in the zone of the nearest filament (see  Appendix \ref{appendix}) do not yet have reliable distance estimates. Obviously, 
this gap needs to be filled in order to better understand the 
structure of the cosmic filament closest to us.

\begin{acknowledgements}
We are grateful to the anonymous referee for a prompt report that helped us
improve the manuscript. This work has made use of the DESI Legacy Imaging Surveys\footnote{\url{https://www.legacysurvey.org/}}, the Sloan Digital Sky Survey (SDSS)\footnote{\url{https://www.sdss3.org/}}, the NASA/IPAC Extragalactic Database\footnote{\url{http://www.ned.ipac.caltech.edu}} (NED), The Galaxy Evolution Explorer (GALEX)\footnote{\url{http://www.galex.caltech.edu/index.html}}, HyperLeda\footnote{\url{http://leda.univ-lyon1.fr/}}, and the revised version of the Local Volume galaxy database\footnote{\url{http://www.sao.ru/lv/lvgdb}}. The Local Volume galaxies database has been updated within the framework of grant 075--15--2022--262 (13.MNPMU.21.0003) of the Ministry of Science and Higher Education of the Russian Federation.
\end{acknowledgements}

\clearpage

\onecolumn
\begin{appendix}
 \section{M101 filament of galaxies }
 \setcounter{table}{0}
\begin{longtable}{lccrrccc}
 \caption{\label{appendix}The list of 72 galaxies in the inspected region with 
distance estimates from 5 Mpc to 10 Mpc} \\
\hline
Name                 &  SGL   &     SGB   &$\log L_K$& T &$V_{\rm LG}$& $D$   &   meth  \\ 
  \hline
 (1)                 &  (2)  &    (3)   &      (4)  &   (5)  &    (6)  &    (7)  &    (8)\\
 \hline
  \endfirsthead

\caption{continued.}\\
\hline
 Name               &  SGL   &     SGB   &$\log L_K$& T &$V_{\rm LG}$& $D$   &   meth  \\ 
  \hline
(1)                 &  (2)  &    (3)   &      (4)  &   (5)  &    (6)  &    (7)  &    (8)\\
 \hline
 \endhead
\hline
\endfoot
\hline
\endlastfoot
\hline
 {\bf M 63}            &  76.20 &  14.25    & 11.00    & 4 &    562     &9.04   &   TRGB \\
 KK 194                &  73.82 &  14.85    &  7.15    & 10&     -      &9.04   &   mem  \\
 Dw1311+4317           &  74.80 &  13.70    &  7.21    & -2&      -     &9.04   &   mem  \\
 Dw1315+4304           &  75.10 &  14.25    &  7.13    & -2&      -     &9.04   &   mem  \\
 dw1303+42             &  75.46 &  12.00    &  6.88    & 10&     -      &9.04   &   mem  \\
 dw1321+4226           &  75.94 &  15.24    &  6.50    & -2&      -     &7.30   &   sbf \\ 
 UGC 8313              &  75.96 &  13.92    &  8.44    & 8 &    683     &8.77   &   sbf  \\
 dw1305+41             &  76.02 &  12.33    &  7.28    & 10&     -      &9.04   &   mem  \\
 KK 191                &  76.12 &  13.85    &  6.83    &10 &     -      &8.28   &   TRGB \\
 TBGdw1                &  76.15 &  13.59    &  6.20    & 10&      -     &9.26   &   sbf  \\
 dw1310+4153           &  76.19 &  13.28    &  6.90    & -1&            &7.38   &   sbf  \\
 KKH 82                &  76.36 &  13.69    &  7.77    & 10&    588     &7.58   &   sbf  \\
 TBGdw5                &  76.63 &  13.91    &  6.52    &-2 &     -      &9.04   &   mem  \\
 KK 193                &  76.73 &  14.11    &  7.03    &10 &     -      &9.65   &   sbf  \\
 dw1315+4123           &  76.83 &  14.05    &  6.43    & -2&      -     &8.80   &   sbf  \\
 dw1308+40             &  77.13 &  12.76    &  7.54    & -1&      -     &9.04   &   mem  \\
 Dw1311+4051           &  77.26 &  13.30    &  6.55    & 10&      -     &9.04   &   mem  \\
 CGCG 217-018          &  77.62 &  13.47    &  8.19    & 9 &    608     &8.99   &   TRGB \\
 DDO 182               &  79.27 &  18.99    &  7.89    & 10&     730    &8.90   &   TRGB \\
 dw1305+38             &  79.85 &  11.73    &  7.03    & 10&     -      &9.04   &   mem  \\
\hline
 {\bf M 51}            &  71.12 &  17.33    & 10.97    &  5&     538    &8.40   &   SN   \\
 dw1327+51             &  66.44 &  17.19    &  7.00    & -1&       -    &8.40   &   mem  \\
 dw1338+50             &  68.29 &  18.96    &  7.08    & -1&       -    &8.40   &   mem  \\
 UGCA 361              &  68.44 &  17.96    &  8.13    & -1&       -    &8.40   &   mem  \\
 LV J1328+4937         &  68.61 &  17.28    &  7.12    & 10&     497    &7.73   &   TRGB \\
 KK 206                &  69.21 &  18.04    &  8.25    &  9&      690   &9.31   &   TF   \\
 LV J1342+4840         &  69.72 &  19.49    &  7.60    &  9&     543    &8.40   &   mem  \\
 NGC 5229              &  70.46 &  18.08    &  8.56    &  7&     456    &8.95   &   TRGB \\
 NGC 5195              &  71.09 &  17.35    & 10.59    &  0&     548    &7.66   &   sbf  \\
 dw1328+4703           &  71.28 &  17.06    &  6.51    & -2&        -   &8.35   &   sbf  \\
 dw1313+46             &  71.45 &  14.40    &  6.98    & 10&       -    &8.40   &   mem  \\
 dw1340+45             &  72.87 &  19.07    &  6.74    & 10&      -     &8.40   &   mem  \\
 MCG+08-25-028         &  73.98 &  18.31    &  7.78    & 10&     565    &8.47   &   TRGB \\
 PGC 2229803           &  74.65 &  16.66    &  7.42    &  9&      529   &6.58   &   NAM  \\
 Dw 1341+42            &  76.72 &  19.08    &  5.93    &  10&       -   &8.40   &   mem  \\
 \hline                                                                                    
 {\bf M 101}           &  63.58 &  22.61    & 10.79    &  6 &     378   &6.95   &   TRGB \\
 Dw1348+60             &  57.61 &  20.18    &  6.75    &  10&      -    &6.95   &   mem  \\
 dw1446+58             &  57.83 &  27.82    &  6.49    &  10&      -    &6.95   &   mem \\
 dw1416+57             &  59.49 &  24.10    &  6.95    &  -2&      -    &6.95   &   mem \\
 dw1343+58             &  59.59 &  19.62    &  7.46    &   9&     338   &5.59   &   TRGB \\
 DDO 194               &  59.68 &  26.65    &  8.09    &   8&     381   &5.81   &   TRGB \\
 NGC 5585              &  60.70 &  24.67    &  9.21    &   7&     457   &7.00   &   TRGB \\
 KK 207                &  61.43 &  18.32    &  6.47    &  10&      -    &6.95   &   mem  \\
 dw1412+56             &  61.50 &  23.70    &  6.79    &  -2&      -    &6.95   &   mem \\
 M101Dw7               &  62.75 &  23.15    &  6.14    &  -2&       -   &6.95   &   mem \\
 M101Dw9               &  62.80 &  21.48    &  6.26    &  -2&       -   &7.73   &   TRGB \\
 dw1350+5441           &  63.32 &  20.82    &  7.54    &  -2&      -    &6.27   &   sbf  \\
 GBT1355+54            &  63.34 &  21.38    &  7.10    &  11&     345   &6.95   &   mem  \\
 NGC 5477              &  63.43 &  22.94    &  8.26    &   9&     451   &6.76   &   TRGB \\
 M101-df2              &  63.53 &  23.40    &  6.74    &  -2&       -   &6.87   &   TRGB \\
 UGC 8882              &  63.91 &  21.76    &  7.51    &  -1&     482   &6.95   &   sbf  \\
 M101-df1              &  64.01 &  22.72    &  6.17    &  10&       -   &6.37   &   TRGB \\
 NGC 5474              &  64.30 &  22.93    &  9.21    &   8&     424   &6.98   &   TRGB \\
 M101-df3              &  64.38 &  22.65    &  7.27    &  -2&       -   &6.52   &   TRGB \\
 M101-DwA              &  64.19 &  23.19    &  6.68    &  -2&       -   &6.65   &   TRGB \\
 Holm IV               &  64.14 &  21.40    &  8.62    &   8&     272   &7.24   &   TRGB \\
 Dw1358+52             &  65.18 &  21.95    &  6.52    &  -2&       -   &6.95   &   mem \\
 dw1355+51             &  66.28 &  21.54    &  7.07    &  -2&      -    &6.95   &   mem  \\
 Dw1409+51             &  66.89 &  23.76    &  6.38    &  10&      -    &6.95   &   mem \\
 Dw1351+50             &  68.07 &  21.08    &  5.75    &  10&      -    &6.95   &   mem  \\
\hline
 {\bf NGC 6503}        &  33.14 &  34.63    & 10.00    &  6 &     309   &6.25   &   TRGB\\ 
 KK 242                &  32.81 &  34.73    &  6.47    &  10&     179   &6.46   &   TRGB \\
\hline                                                                                    
 UGC 7748              &  48.59 &  12.71    &  7.68    &   9&     590   &8.70   &   TF   \\
 Dw1245+6158           &  55.23 &  12.76    &  6.41    &  10&      -    &5.60   &   mem  \\
 NGC 4605              &  55.47 &  12.02    &  9.70    &   8&     284   &5.55   &   TRGB \\
 KDG 162               &  58.56 &  10.76    &  6.87    &  10&      -    &10.0   &   txt  \\
 KKH 87                &  60.42 &  23.98    &  7.73    &  10&     473   &8.87   &   TRGB \\
 KDG 192               &  63.23 &  11.02    &  9.66    &  10&     544   &9.66   &   TRGB \\
 UGC 7950              &  65.67 &  11.02    &  8.36    &   9&     599   &8.90   &   TF   \\
 NGC 4707              &  66.16 &  11.15    &  8.25    &  10&     588   &6.52   &   TRGB \\
 BTS 157               &  70.34 &  10.80    &  7.22    &  10&     644   &9.04   &   mem  \\
 NGC 5023              &  74.05 &  13.89    &  6.05    &   7&     476   &6.05   &   TRGB \\
\hline                                                                                    
\end{longtable} 
\end{appendix}

\end{document}